# Auger spectroscopy of strongly correlated systems: present status and future trends

Claudio Verdozzi\*\*, Michele Cini\* and Andrea Marini\*

\*  Istituto Nazionale per la Fisica della Materia and Dipartimento di Fisica, Università di Roma "Tor Vergata", Via della Ricerca Scientifica 1, I-00133 Roma, Italy
\*\* Department of Physics and Astronomy , University of Edinburgh, JCMB, Mayfield Road, Edinburgh EH9 3JZ Scotland UK

We review the Cini-Sawatzky approach to the line shape analysis of Auger *CVV* transitions, in its basic and original formulation. Then, several extensions of the theory are reviewed, namely the inclusion spin-orbit coupling, dynamical/plasmon screening, overlap effects, off-site interactions, the treatment of disorder and the formulation for partially filled bands. We conclude by highlighting what we consider to be future directions in the field.

## 1. INTRODUCTION

In the atomic Auger decay [1, 2], a deep electron vacancy (hole) recombines with an outer electron, with a second electron ejected in the energy continuum. The information potential of this process is probed by Auger Electron Spectroscopy (AES). Compared to atomic ones, solid state Auger transitions may introduce qualitatively new features since core levels are to a good approximation the same as in isolated atoms, but valence ones are significantly different. So, transitions involving two valence holes in the final state, so called Core-Valence-Valence (CVV) transitions, render the theoretical description more difficult than in atoms since the valence holes may exhibit a competition between itinerant and localized behavior. The presence of a localized (core) level in the CVV Auger matrix element means that the lineshape should actually give information about the *local* distribution of valence levels, specifically about the two-hole local density of states, 2LDOS. This information can be most efficiently extracted if we can compare experimental data with calculated electronic properties, e.g. band structure calculations, theoretical density of states , etc. In the earliest model of Auger line shapes developed by Lander [3] within the band theory of solids, the Auger spectrum is proportional to the self-convolution of the density of occupied states. However, Powell [4] pointed out that experimentally this description was qualitatively good in some cases but totally failed in other instances, when the spectrum presented sharp *quasiatomic* features. Phenomenological models were proposed [5] to describe the experimental line shapes; eventually they were superseded by the similar approaches of Cini [6] and Sawatzky [7] which provided a satisfactory explanation of the experimental CVV Auger Spectra of transition metals with closed valence bands [8]. The CS model/theory (from now onwards referred to as CSM or CST) allows to understand the phenomenology involving band-like, atomic-like and intermediate situations in terms of the U/W ratio of the on-site repulsion U to the band width W. In both models, the spectrum was expressed in terms of the interacting local two-holes Green's function $G(\omega)$, which could be found exactly. The quasiatomic structures were interpreted as two-hole resonances, i.e. poles of $G(\omega)$. For intermediate U/W, atomic-like peaks and band-like structures are predicted and observed [8] in the same spectrum. For high U/W (atomic-like case) the Auger line shapes are so close to the free-atom spectra that they are labeled by LSJ terms and levels. Later, the two-hole resonances were found to be important for understanding the stimulated desorption as well, in the Knotek-Feibelman mechanism [9].This intuitive approach proved to be flexible enough to allow the extensions needed for the comparison with experiment in a variety of situations. Plasmon satellites, off-site interactions, interatomic overlaps, and disorder effects, which were outside the original scope, were successfully included, gaining clear insight in the experiments. The extension to incompletely filled bands is difficult, but important partial success has definitely been achieved. While *ab initio* calculations are being developed for small molecules [10] (see however [11] for a specific case from solid state) we wish to outline here the development of the present, semiempirical approach which is valuable in complex/ novel situations (expecially where qualitative issues still need to be addressed), thus paving the way to more quantitative treatments.

## 2. CSM : THE BASIC VERSION

A treatment of correlations for closed bands was introduced in [6] in terms of an Anderson Hamiltonian; then, a Hubbard Hamiltonian approach was proposed in [7]. The relation between the two approaches was clarified in [12,13]. Besides matrix elements, the spectrum depends on $D(\omega)$, the two-hole *local* density of states (2LDOS). Let $|0\uparrow 0\downarrow\rangle$ be the state of two holes with opposite spin at the site where the Auger transition occours ( the *Auger site*, for short ). In the Cini model,

$D(\omega) = \langle 0\uparrow 0\downarrow | \delta(\omega-H) | 0\uparrow 0\downarrow\rangle$

$= \sum_\lambda |\langle\lambda | 0\uparrow 0\downarrow\rangle|^2 \delta(\omega - E_\lambda) = \frac{1}{\pi}\operatorname{Im} G(\omega)$,  (2.1)

$G(\omega) = \langle 0\uparrow 0\downarrow | \frac{1}{\omega\text{-H-}i0^+} | 0\uparrow 0\downarrow\rangle$   (2..2)

with $H = H_0 + U n_{0\uparrow}n_{0\downarrow}$, $H_0$ the band term, and $|\lambda\rangle$ the two-hole eigenstates of H. One finds

$G(\omega) = \frac{G_0(\omega)}{1 - UG_0(\omega)}$,   (2.3)

with

$G_0(\omega) = \int_{-\infty}^{\infty} \frac{N(\omega')}{\omega - \omega' - i0^+} d\omega'$,

$N(\omega)$ is the self-convolution of the non interacting one-body Local Density of States (1LDOS), and carries the band structure information. The key quantity, fully determining the transition from band-like (U~0) to atomic like regimes is the ratio U/W (Figure 1) , with W the bandwidth. For low U/W, the line shape is close to the self-convolution of the local one-hole density of states; on increasing U/W, the shape is distorted until for a critical value of the ratio, two-hole resonances appear, that correspond to poles of $G(\omega)$ . For U/W exceeding a critical value, a pole occurs in $G(\omega)$ and a resonant state develops outside the band continuum.

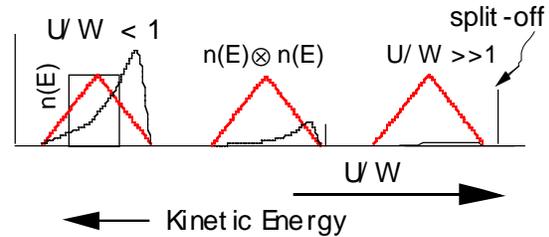

**Figure 1** Dependence o the lineshape on U/W

For higher U/W values most of the intensity is in the resonant (split-off) state at the expense of the band region. In real systems, the resonances have a finite width, due to various decay mechanisms [14] . For a periodic system in the Hubbard/Sawatzky formulation, the resonance also acquires a dispersion width since the holes hop together as an exciton-like state. The split-off states, spatially localized on the site of the Auger decay are the sharp features seen in atomic-like spectra. In comparisons with the experiment, one needs to include band degeneracy [6] and, for heavier elements, the Spin-Orbit Interaction [15]. Then, in the Cini model, H becomes

$H = H_0 + \sum_{SLJM} U(SL) |SLJM\rangle\langle SLJM|$   (2.4)

$H_0$ contains j-resolved ($\vec{J} = \vec{L} + \vec{S}$) band structure information, the U(SL) for d-bands are given by $F^0$, $F^2$, $F^4$ Slater's integrals (in solids, $F^0$ is largely reduced but $F^2$, $F^4$ keep approximately their atomic values [16]) and $|SLJM\rangle$ is the two hole state at the Auger site. In the intermediate coupling (IC) scheme, different total J channels decouple, and one gets a matrix problem for each J, mixing only the

appropriate SL components. The Auger spectrum $A(\omega)$ is then given by [15]

$$A(\omega) = \sum_J \sum_{\mu\nu} M^*_\mu(J) M_\nu(J) \frac{1}{\pi} \text{Im}[\mathbf{G}^J(\omega)]_{\mu\nu} \quad (2.5)$$

while $M_\mu(J)$ and $|\mu\rangle \equiv |\mu(J)\rangle$ are atomic IC transition rates and eigenvectors, respectively, and $[\mathbf{G}^J(\omega)]_{\mu\nu}$ is the matrix element of the 2-hole Green's operator; since

$$|\mu(J)\rangle = \sum_{SLJ} \langle SLJ | \mu(J)\rangle | SLJ\rangle \quad (2.6)$$

$\mathbf{G}^{(J)}(\omega)$ is obtained by a matrix inversion in SLJM basis (**1** is the unity matrix)

$$\mathbf{G}^{(J)}(\omega) = \frac{\mathbf{1}}{\mathbf{1} - \mathbf{G}_0^{(J)}(\omega)\mathbf{U}} \mathbf{G}_0^{(J)}(\omega), \quad (2.7)$$

$$\langle SLJ | \mathbf{U} | S'L'J'\rangle = U(SL)\delta_{SS'}\delta_{LL'}\delta_{JJ'} \quad (2.8)$$

the $\mathbf{G}_0^{(J)}(\omega)$ matrix is related via the 9-j symbols to the convolution of the 1LDOS in jj coupling.

## 3. DYNAMICAL SCREENING EFFECTS

The Auger spectra, like the photoemission ones, commonly show plasmon satellites and asymmetric loss features due to electron-hole pair excitations; these are known as final-state effects. Electron-hole excitations can be bosonized and treated as oscillators like the plasmons. The relative intensity of the first plasmon satellite is clearly related to the probability that the Auger electron leaves the solid with one plasmon excited. The excitation is due to the coupling of the plasmons to the local charge at the Auger site, which is suddenly changed by the transition. The valence holes, in turn, feel the potential of the boson modes in the system and adjust their motions accordingly; therefore, to properly understand the line shape, one must consider the coupled system of holes and plasmons relaxing together. As a results, the plasmon satellites are not rescaled, shifted replicas of the main line, and the main line is deformed by the dynamical interaction with the plasmons. Also, in most the Auger literature, U is a screened repulsion, corresponding to the limit of static screening: in a broader contest, dynamical screening effects would naturally appear in a One Step Model treatment, (see Sect.7 below). In a Two-Step-Model, TSM scheme, the effect of plasmons on the lineshape was studied in [17], where it was shown that if one describes the plasmon dynamics by one effective mode, the generalised Auger model is solved exactly for closed bands in terms of the method of Excitation Amplitudes [18], a continued fraction technique. The extended model Hamiltonian is

$$H_{EH} = H_b + H' \quad (3.1)$$

including a *bare* term (no plasmons)

$$H_b = H_{TB} + H_I$$
$$H_I = U_0 \sum_R n^r_{R\uparrow} n^r_{R\downarrow} \quad (3.2)$$

where $H_{TB}$ is a tight-binding model; the plasmons and their interactions are contained in

$$H' = H_p + H_{h-p}$$
$$H_p = \sum_q \omega_q^r b_q^{r\dagger} b_q^r \quad (3.3)$$

where

$$H_{h-p} = \sum_q g_q^r \sum_{R\sigma} e^{i\vec{q}\cdot\vec{R}} n^r_{R\sigma} b_q^\dagger + \text{h.c.} \quad (3.4)$$

With the two holes regarded as a double charge when $R=R'$ (and as two single charges for $R \neq R'$) and for a large upper cutoff $q_c$ in the q-sums in Eq.(3.3, 3.4) the relaxation shift is

$$\Delta E(\mathbf{R}\mathbf{R'}) = 2(1+\delta_{\mathbf{R}\mathbf{R'}})\sum_q \frac{g_q^2}{\omega_q} \quad (3.5)$$

consistently with the core-limit of the theory.
In [17] a diagrammatic analysis was performed for the Auger propagator, starting from a fully relaxed core-hole state for the Auger decay (this is the content of the TSM, which is correct for a long-lived core hole). The full diagram expansion is rather involved; however, in practice $\omega_p > W$, where W is several times the hopping parameter $V$, so keeping only the leading diagrams in $\lambda = V/\omega_p$ and neglecting dispersion ($\omega_q \equiv \omega_p$) the result coincides with one where the plasmon field is represented by a single effective mode, and the Hamiltonian becomes:

$$H_{OB} = [H_{TB} + U_0 P] + \omega_p b^\dagger b +$$
$$[2g_0 P + g_0\sqrt{2}(1-P)](b^\dagger + b), \quad (3.6)$$

where $P = \sum_R n_{R\uparrow}^r n_{R\downarrow}^r$, $g_0^2 = \sum_q g_q^2$. One obtains the *static* (fast-plasmon, or narrow-band) limit of the theory by averaging the boson operators on the fully relaxed coherent state for the instantaneous two hole configuration. In this limit, we get an effective hamiltonian

$$H_{\text{eff}} = H_{tb} + UP - \delta, \quad \delta = 2g_0^2/\omega_p. \quad (3.7)$$

However, in general we must deal with dynamical screening and cope with (3.6). To this end, we introduce the operator $X(\mu) = \exp[\mu g_0 (b - b^\dagger)]$, which helps to write down the exact local propagator

$$G(\omega) = \exp(-\gamma^2/\omega_p^2)$$
$$\times \sum_{m=0}^{\infty} \sum_{n=0}^{\infty} \frac{(\gamma/\omega_p)^{m+n}}{m!n!} \phi_{nm}(\mathbf{0},\omega), \quad (3.8)$$

in terms of the *excitation amplitudes*

$$\phi_{nm}(\mathbf{R},\omega) = \langle 00, v | b^m \frac{i}{\hat{\omega} - \tilde{H}} (b^+)^n | \mathbf{RR}, v \rangle,$$

$$\phi_{nm}^0(\mathbf{R},\omega) = \langle 00, v | b^m \frac{i}{\hat{\omega} - \tilde{H}_b - H_p} (b^+)^n | \mathbf{RR}, v \rangle$$

$$\phi_{nm}^0(\mathbf{R},\omega) = \delta_{nm} m! \Gamma^0(\mathbf{R}, \omega - n\omega_p),$$

$$\Gamma^0(\mathbf{R},\omega) = \langle 00, v | \frac{i}{\hat{\omega} - \tilde{H}_b} | \mathbf{RR}, v \rangle \quad (3.9)$$

here we used the shorthand notations

$$\tilde{H} = X^+(\sqrt{2}) H_{OB} X(\sqrt{2}) = \tilde{H}_b + H_p + \tilde{H}_{h-p}$$
$$\tilde{H}_b = H_{tb} + U_d P - \delta,$$
$$U_d = U_0 - 4(\sqrt{2} - 1)g_0^2/\omega_p$$
$$\tilde{H}_{h-p} = gP(b + b^+),$$
$$\delta = -2g_0^2/\omega_p, \ g=(2-\sqrt{2})g_0, \ \gamma=g_0(\sqrt{2}-1).$$

The *excitation amplitudes* are first obtained in Fourier space; i.e. one starts with

$$\phi_{nm}(\mathbf{0},\omega) = \frac{1}{N} \sum_{\mathbf{k}} \phi_{nm}(\mathbf{k},\omega), \quad (3.10)$$

and uses the identities

$$\frac{1}{z-A-B} = \frac{1}{z-A} + \frac{1}{z-A} B \frac{1}{z-A-B}, \quad (3.11)$$
$$b^m(\tilde{H}_b + H_p) = (\tilde{H}_b + H_p + m\omega_p)b^m.$$

Then after setting $\phi_{nm}(\mathbf{k},\omega) = \phi_{nm}$,

$$\Gamma_n^0 = \Gamma^0(\mathbf{k}, \omega - n\omega_p), \quad (3.12)$$

one finally gets, after some algebra,

$$\phi_{nm} = \Gamma_n^0 [\delta_{nm} n! - ig(n\phi_{n-1m} + \phi_{n+1m})].$$

Solving this, one obtains the off-diagonal amplitudes in terms of the diagonal ones:

$$\phi_{nm} = \begin{cases} A_{n+1} A_{n+2} \ldots A_m \phi_{mm} & n < m \\ \\ B_{n-1} B_{n-2} \ldots B_m \phi_{mm} & n > m \end{cases} \quad (3.13)$$

with

$$A_n = \cfrac{-ig\Gamma_{n-1}^0}{1 + \cfrac{g^2(n-1)\Gamma_{n-1}^0 \Gamma_{n-2}^0}{1 + \cfrac{g^2(n-2)\Gamma_{n-2}^0 \Gamma_{n-3}^0}{1 + \ldots}}}$$

$$B_n = \cfrac{-ig(n+1)\Gamma_{n+1}^0}{1 + \cfrac{g^2(n+2)\Gamma_{n+1}^0 \Gamma_{n+2}^0}{1 + \cfrac{g^2(n+3)\Gamma_{n+2}^0 \Gamma_{n+3}^0}{1 + \ldots}}}$$

Eventually, the diagonal terms are obtained:

$$\phi_{mm} = \frac{m! \Gamma_m^0}{1 + ig\Gamma_m^0 (mA_m + B_m)} \quad (3.14)$$

This "dynamical" treatment may have a non-trivial effect on the lineshape due to the additional energy channel now available: variations of the plasma frequency $\omega_p$, the interaction $U_d$ and bandwidth W, allow the two-hole resonances to have extra broadening and possibly delocalize. The formulation also fixes the range of applicability of the "static" CST, and has been applied to the Auger spectrum of Graphite [19], and Ag [20], generalizing the method to degenerate bands.; g and $\omega_p$ are obtained from experiment, so no new free parameters are involved.

## 4. OFF-SITE INTERACTIONS

A basic feature of the CS theory (CST) is that U fixes the energetics of the spectrum and its line-shape simultaneously, and that U/W is the only characteristic ratio. However, high resolution CVV spectra of Au [21] and Ag [20] showed that using U as an adjustable quantity, an optimization of the theoretical line-shape resulted in poor energy alignment with experiment and vice-versa. This behavior was understood in terms of the Off-Site Interactions (OSI), which are neglected in the CST [22,23]. The concept that the OSI may affect the line-shape was first discussed in [24] for molecules; similar ideas were later applied to solids [25] on a semiempirical basis. We reproduce here our <u>exact</u> treatment of the OSI in the Extended Hubbard Model EHM [26, 27]. For a general interaction $U(r)$, the Hamiltonian reads

$$H = \sum_{k,s} E(\vec{k}) n^r_{k,s} + \sum_{R,r} U(r) n^r_{R\uparrow} n^r_{R+r\downarrow} \quad (4.1)$$

$$n^r_{k,s} = a^\dagger_{ks} a^r_{ks}, \; n^r_{R,s} = a^\dagger_{R,s} a^r_{R,s} \quad (4.2)$$

and $E(\mathbf{k})$ are the Bloch energies of a non-degenerate band. If $|\mathbf{R},\mathbf{R+r}\rangle$ is the ket with the up spin hole at $\mathbf{R}$ and the down spin hole at $\mathbf{R+r}$, we can go to a mixed picture $|\mathbf{Qr}\rangle$ where $\mathbf{Q}$ is the center-of-mass wavevector, via the transformation

$$|\mathbf{R,R+r}\rangle = N^{-1/2} \sum_Q e^{-i\mathbf{Q}(\mathbf{R}+\frac{\mathbf{r}}{2})} |\mathbf{Qr}\rangle, \quad (4.3)$$

$$|\mathbf{Qr}\rangle = e^{i\frac{\mathbf{Qr}}{2}} (N)^{-1/2} \sum_q e^{-i\mathbf{qr}} |\mathbf{Q-q, q}\rangle \quad (4.4)$$

In the $|\mathbf{Qr}\rangle$ basis, H assumes the form:

$$H = \sum_\mathbf{Q} H^\mathbf{Q}, \quad H^\mathbf{Q} = H_0^\mathbf{Q} + H_1^\mathbf{Q} \quad (4.5)$$

By translational symmetry, the most general two-particle Green's function can be written as

$$\langle \mathbf{0\; 0+r} | \frac{1}{z-H} | \mathbf{R\; R+r'}\rangle =$$

$$\frac{1}{N} \sum_\mathbf{Q} e^{-i\mathbf{QR}} \; e^{i\frac{\mathbf{Q(r-r')}}{2}} G^\mathbf{Q}_{\mathbf{rr'}}(z) \quad , \quad (4.6)$$

$$G^\mathbf{Q}_{\mathbf{rr'}}(z) = \langle \mathbf{Q\; r} | \frac{1}{z-H^\mathbf{Q}} | \mathbf{Q\; r'}\rangle$$

$$= G^\mathbf{Q}_{\mathbf{rr'}}(z) = g^\mathbf{Q}_{\mathbf{rr'}}(z) + \Lambda^\mathbf{Q}_{\mathbf{rr'}}(z) \quad , \quad (4.7)$$

where $g^\mathbf{Q}_{\mathbf{rr'}}$ is given by Eq.(4.6) for $H_1^\mathbf{Q} = 0$, and

$$\Lambda^\mathbf{Q}_{\mathbf{rr'}}(z) = \sum_{ij} g^\mathbf{Q}_{\mathbf{ri}}(z) \left( \frac{\mathbb{1}}{\mathbb{1}- \mathbf{M}^\mathbf{Q}(z)} \right)_{ij} U(j) \; g^\mathbf{Q}_{\mathbf{jr'}}(z) \; ,$$

$$[\mathbf{M}^\mathbf{Q}(z)]_{\mathbf{rr'}} = U(r) g^\mathbf{Q}_{\mathbf{rr'}}(z) \quad (4.8)$$

The approach, rephrased in terms of a non-Hermitean effective Hamiltonian, amounts to solve the two-holes dynamics in an infinite system by a matrix inversion in an "interaction cluster", whose size depends on the range of the potential. For two holes with parallel spins, the lineshape depends only on $U(r\neq 0)$. One qualitatively new feature compared to the CST is that a general $U(r)$, produces several resonances in the 2LDOS instead of one; this reflects the fact that long ranged potentials may have many bound states as illustrated in Fig. 2 (adapted from [27]), where the interaction values are truncated after 2 lattice parameters (the nonzero values are shown in the Fig.2).

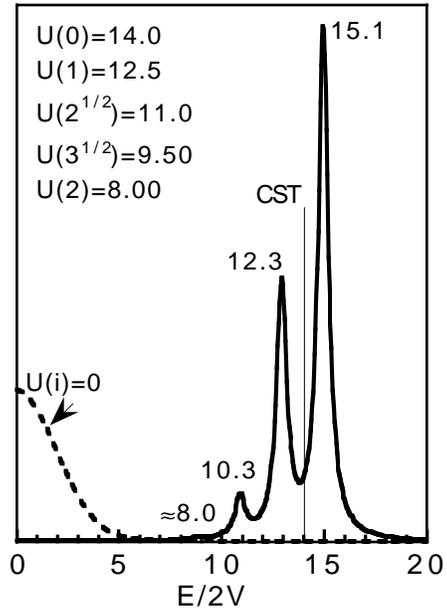

**Figure 2** 2LDOS results or the OSI-EHM (see text); noninteracting limit (dashed); an example with a rather strong interaction producing several resonances (solid).

The dashed line is the non-interacting line shape (2LDOS band-edge=6V, V=hopping parameter). In the interacting case (solid) all the structures are split-off; the delta function in the simpler CST solution for no OSI is also shown (bar). In travelling the crystal, the particles are delocalized within a distance r ≤ 2a. To qualitatively highlight the effect of OSI in Auger spectra, in Ref. [26] an indipendent multiplet analysis was performed for Ag and Au, explicitly considering $^1S_0$ (the most atomic-like) $^1G_4$ (the most intense) terms, using a potential U(r) with a Thomas-Fermi form U(r)= U(0) $\delta_{r0}$+θ(0<r≤2a) (A/r) $e^{-\lambda r}$. For both Ag and Au, the search of the best CST fit gave consistent values of the energy shifts for the two LSJ terms. The overall effect of OSI was a) an energy shift between the two particle local density of states as calculated in the OSI theory and the CST Auger lineshape b) the lineshape was only affected to a minor extent, which is consistent with the CST shapes already agreeing well with experiment. This showed that the EHM extension of CST provides a more general framework, expected to be especially relevant in system where screening is not very effective (ionic compounds, oxides, disordered systems) or can be varied in a ontrolled way [28]. Considerations about a multiple-band OSI formalism for comparison with experiment, and a review of early use of OSI in Auger spectroscopy, can be found in [27].

**5. OVERLAP EFFECTS**

Early theoretical work on the Si $L_{23}$VV line shape neglected the hole-hole interaction and was based on semi-empirical tight binding (SETB) calculation of the local DOS [29-31]. The experimental spectrum was clearly band-like, and consisted of *ss*, *sp* and *pp* regions; the non-interacting theories gave a rough overall agreement, except that they predicted an excessive *sp* intensity at about 15 eV binding energy. The SETB density of states can also be used to set up the CST line shape, but one finds that including interaction effects does not improve the situation, and the exaggerated build-up of *sp* intensity remains there. Since the CST works well for transition metals, we argued [32-33] that the difficulty with covalently bonded solids [34] must arise from its neglect of the overlap between atomic orbitals $\phi_{R,\Lambda}^{\mathbf{r}}$ at different sites $\vec{R}$; Λ stands for the set of angular momentum quantum numbers. Accordingly, we proposed and tested an extended model, which was designed to remain as close as possible to the original CST. Accordingly, if $\vec{R}=0$ is the site of the Auger decay, we must continue using the orbitals $\phi_{0L}$ to compute the Auger and U matrix elements, and the projected DOS matrix is also defined in terms of $\phi_{0L}$. The need to generalize arises in that the tight binding Hamiltonian refers to an <u>orthogonal</u> basis set; this set must be the basis of the same representation of the group of the Schrödinger equation as the atomic orbitals (AOREP). Such a basis is provided by the Löwdin orbitals [35] arranging the atomic orbitals in a row vector $\Phi(\mathbf{r})$ we have

$$\Psi_L(\mathbf{r}) = \Phi(\mathbf{r})\mathbf{S}^{-1/2}, \qquad (5.1)$$

where **S** is the atomic overlap matrix. This basis is not uniquely determined; in particular, we may consider any unitary matrix **V** commuting with the AOREP, and build the alternative Löwdin set

$$\Xi_L = \Psi_L V; \qquad (5.2)$$

it is clear that the transformation **V** affects the Hamiltonian in such a way that its eigenvalues and symmetry properties remain unaltered. The tight binding model Hamiltonian **H** is parametrized in such a way to fit the energy bands of the solid as closely as possible, but this gives us no information at all about **V**. In other terms, an unknown matrix **V** relates the basis set used to calculate the Auger and the screened Coulomb matrix elements to the one used to calculate the local DOS. This matrix, however, can be modeled and *observed* by fitting the experimental spectrum. We took the matrix **V** such that each Löwdin orbital is multiplied by a different phase factor of the form $\exp[i\alpha]$, where $\alpha$ is a real number.

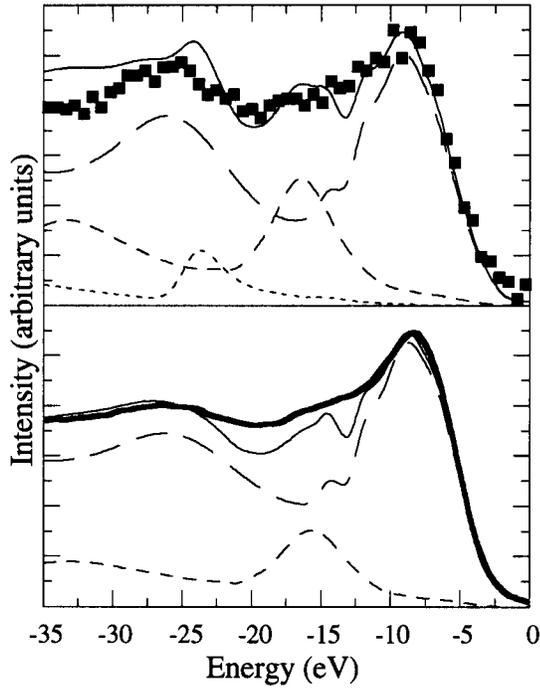

**Figure 3** Experimental Si-$KVV$ (dot) and Si-$L_{23}VV$ (thick) line shapes compared to the extended CST (thin) with $U_s^s = U_p^s = 6$ eV and the atomic DOS with the phase $\alpha = 3\pi/4$. The $pp$ (long dashed), $sp$ (medium dashed) and $ss$ (short dashed) components are also shown. To allow the comparison between the theoretical and the experimental line shape on an absolute energy scale, the core-hole binding energy relative to the Fermi level, $\varepsilon_{L_{23}} = 99.7$ eV and $\varepsilon_K = 1839$ eV, was subtracted from the measured Auger electron kinetic energy. All the theoretical and the experimental Auger spectra have been normalized dividing the intensity by the maximum value.

In the case of the valence shell of silicon, the phase factor between $s$ and $p$ orbitals is to be determined. We obtained a continuous set of parameterisations for **H** multiplying all the $sp$ SETB parameters of Papaconstantopoulos-Economou, PE [36], by the factor $\exp[i\alpha]$. We performed a comparative analysis of both the Si-$L_{23}VV$ and Si-$KVV$ Auger experimental spectra versus $\alpha$. The atomic multiplets were considered as is usual in the CST but the weak spin-orbit coupling was neglected for simplicity. The Auger matrix elements and the free-atom Coulomb matrix $\mathbf{U}^0$ were computed from [37]. A statically screened Coulomb matrix $\mathbf{U} = \mathbf{U}^0 - \mathbf{U}^s$ was assumed. $\mathbf{U}^0$ being diagonal, to a very good approximation, in the multiplet state basis set, we assumed the screening matrix $\mathbf{U}^s$ diagonal in the same basis:

$$U_{sspp}^s(^1S) = 0$$
$$U_{ss}^s(^1S) = U_{sp}^s(^1P) = U_{sp}^s(^3P) = U_s^s \qquad (5.3)$$
$$U_{pp}^s(^1S) = U_{pp}^s(^1D) = U_{pp}^s(^3P) = U_p^s.$$

A constant background was subtracted from experimental spectra. Auger electron energy loss and experimental broadening were approximately included in the theoretical spectra, by convoluting with back scattered electron spectra with primary energy 300.1 eV for $L_{23}VV$ and 1991.6 eV for $KVV$ spectrum. In Figure 3 we present our best fits obtained using only one screening parameter $U_s^s = U_p^s = 6$ eV and the phase $\alpha = 3\pi/4$. Introducing the phase dramatically improves the fit. Auger matrix element effect favor the $ss$ and $sp$ components in the $KVV$ more than in the $L_{23}VV$ spectrum. We stress that the inter-atomic phase factors are now measured quantities, that can in principle be observed by several other spectroscopies. They lead to a fuller characterization of the system in the SETB method.

## 6. ALLOYS AND DISORDER

In the mid-80's, a vast amount of important experimental [38] and theoretical work [39] was performed on the Auger spectra of alloy systems. Interpreting experiments of many interesting alloys, for example those made of transition metals, may require the simultaneous account of correlation and disorder. A review of the theory of Auger spectra from disordered / concentrated alloys in terms of the Coherent Potential Approximation (CPA) is in [40-42], whilst we refer to [43-45] for the dilute case. This latter case is very important , since in this category fall systems which may exhibit complex ground-state (notably, Kondo-) behaviour [46]. In [47], a numerical approach to the Auger lineshape for disordered systems was considered, especially suitable for those exhibiting a "conventional" ground state. The model is a single band SC superlattice, r-mesh, and the unit supercell is a cubic array of $n^3=N_d$ sites, d-mesh, populated by A, B atoms. Substitutional disorder was included by selecting a quasi-random supercell, with constraints imposed on concentration, local shell coordination, etc. The method allows "exact" disorder space averaging, over a single configuration $C_0$ ( a consistent procedure in the thermodynamical limit, provided ergodicity holds) for the selected cell and the constrained search optimises the cell size n to get results of significance when $N_d \to \infty$. Also, the approach naturally lends itself to a real-space *ab-initio* approach to the Auger lineshape for disorderd systems, similar to those which have been considered for the Photoemission [48] process. In [47], a single band, disordered, $A_{50}B_{50}$ alloy, Auger two-body problem was considered, tests on cell-size effects performed, and the results are shown in Fig. 4. Alben et al's [49] (Fig.4, inset) 1LDOSs with cells of ~ 10000 sites are commonly retained to be in the self averaging limit. The constrained cell model had $N_d=4096$, and (Fig. 4, inset) performed significantly better than CPA. For the parameters to be used for the Auger simulations below, 1LDOS results from cells with 4096, 512, 216 sites show (Fig. 4) that even the 1LDOS standard deviations for the three cell sizes are in good agreement, and that already a 216 site cell is a minimal-size representative of the $N_d \to \infty$ case, also for LDOS fluctuations. With such a cell, one considers the local, conditionally averaged propagator for two holes created at a specific type of alloy site (in principle, by experimentally selecting the core binding energy). The Hamiltonian used is

$$H = \sum_{rd;r'd'\sigma} V_{dd'}(r-r')a^\dagger_{rd\sigma}a_{r'd'\sigma} + \sum_{rd} U_d n_{rd\uparrow}n_{rd\downarrow}$$

where $V_{dd}(0) = E_A$ or $E_B$, $U_d=U_A$ or $U_B$. For diagonal disorder, $V_{d\neq d'}(r)= V$ for all n.n. hoppings, and zero otherwise. For off diagonal disorder, $V_{d\neq d'}(r)= V_{AA}, V_{AB}, V_{BB}$ for the n.n. pairs AA, AB, BB and zero otherwise. Conditionally averaged interacting (A,B curves) and non interacting (A0,B0 curves) 2LDOS at sites A, B are shown in Fig. 5 together with a schematic description of the approaches (denoted by $\alpha,\beta,\gamma$) used (for notational consistency with all other Sections, $\Phi \equiv G$ should be understood in Fig.5 ). The $\gamma^d_{P=A(B)}$ is 1 if the atom type at site d is A(B), and zero otherwise; so the conditional average of a quantity Q(d) can be written as

$$\langle Q \rangle_P = \frac{\sum_d \gamma^d_P Q(d)}{\sum_d \gamma^d_P}, P = A, B.$$

For $U_P=0$, the self-convolution largely reduces the difference between the exact 2LDOS. For $U_P \neq 0$, two different approximation schemes ($\beta,\gamma$) are considered besides the exact intracell solution $\alpha$. The latter is a local approximation with respect to the supercell lattice (the dynamics is exact within the 216-atom supercell) while $\beta$ is a local approximation at the site level (then worse than $\alpha$) and $\gamma$ (clearly incorrect but computationally very simple and useul for the discussion) uses single particle averaged propagators. In both panels of Fig. 5 α, β, γ agree well, but some discrepancies are left: they are mainly due to the quality of the local approximation performed (α describes well the amount of delocalization the two-holes achieve starting from an Auger/ correlation-induced localised state). This delocalisation is underestimated by the more drastic local approximation in β (site A Fig.5a). The problem also shows in homogeneous systems, notably with off-site interactions. To a minor extent, β and γ differ between themselves as well: a further, small broadening is introduced by the average over A sites in β , not present in γ . Discrepancies at A sites reduce greatly for higher U/W values in Fig.5b. A large disagreement is observed at site B, Fig.5b. (in Fig.5a, site B, the agreement is good).

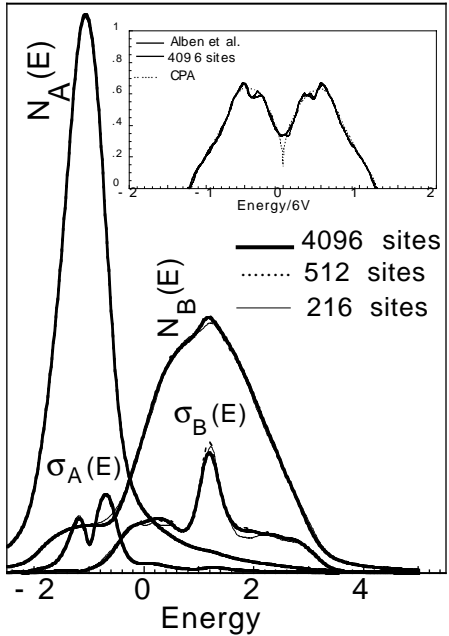

**Figure 4** Conditionally averaged 1LDOS results for different cell sizes (main panel) in the tailored supercell approach. 1LDOSs standard deviations are also shown. A comparison with a CPA treatment and results from are reported in the inset [47].

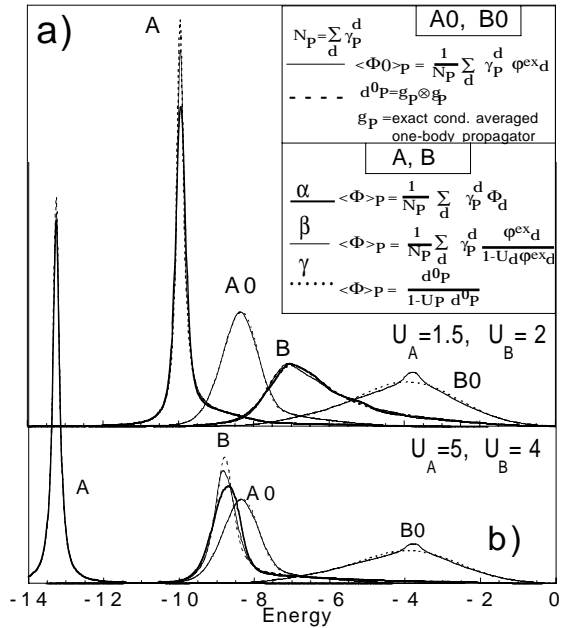

**Figure** 5 2LDOS in a disoredered 216-atom supercell. The same one body parameters (see H in the main text) apply to both panels: $E_A=-1$, $E_B=1$, $V_{AA}=0.15$, $V_{AB}=0.3$, $V_{BB}=0.5$, in arbitrary units.

The explanation involves directly the effect of disorder and alloying, since in homogeneous solids, the local approximation improves for larger U/W ratios. On the contrary, for $U_B$=4, Fig.5b, the agreement is worse than for $U_B$=2, Fig. 5a. Delocalization process like AA (same site) → AB or BB' are not properly accounted in β,γ. This was already noticed in [50] and labeled as "dissociative broadening" and recently re-investigated in [51]. To qualify better the role of disorder in such processes, further investigation are currently being carried, in terms of an improved approach for significantly larger cells (1000-2000 sites).

## 7. OPEN BAND PROBLEM

For open valence bands, no narrow atomic-like peaks exist, since any resonance has plenty of many-body states to decay. Since the system has many degrees of freedom the problem gets hard unless some parameter is small. One relatively simple case is that of almost completely filled bands, when the number of holes per quantum state $n_h \ll 1$ and the closed-band theory can be extrapolated[52]. A diagrammatic analysis shows that correlation and shake-up effects factor out in first-order in $n_h$ ; then we may go ahead by the two-step model, which neglects any coherence effect between creation and decay of the core-hole.

*Two-Step Approach for $n_h \ll 1$*

Thanks to the above mentioned factorization of correlation and shake-up, the XPS and Auger spectra can be computed from the equilibrium one-body and two-body propagators $S(\omega)$ and $G(\omega)$. Galitzkii [53] has shown how to sum the dominant contributions to the diagrammatic expansion in the low density approximation LDA. The theory for $S(\omega)$ led us to the explanation of the high energy satellites seen in valence photoemission from Ni and other transition metals; they arise from two-hole resonances. The dominant diagrams of the perturbation expansion of $G(\omega)$ are just the same ladder diagrams which provide the exact solution for $n_h \rightarrow 0$. This gives a satisfactory explanation of the Auger spectra for $n_h \approx 0.1$, which includes interesting cases like Ni [52-60], Pd bulk [15], Pd in finely dispersed form [61] and Cu/Pd alloys [62]. The case in [61] is very amusing, since one can tune the band filling (and a metal-insulator transition) by the cluster size and observe the changes in self-energy.

To extend the LDA to somewhat larger $n_h$, Drchal and Kudrnovsky [63] worked out the self-consistent version of the LDA that involves using dressed propagators in the ladder series. However cluster studies [57-58] unexpectedly but clearly showed that the ladder approximation with bare propagators is superior and is a good approximation to $G(\omega)$ for $n_h < 0.25$ and a range of U/W. This is the *Bare Ladder Approximation* (BLA), that we wish to examine in some detail here. We introduce a special notation $\alpha l, \beta l, ...$ to denote the set of quantum numbers of the local valence spin-orbitals belonging to the *Auger site*. Ordering the spin-orbitals in an arbitrary way, we may write the Coloumb valence-valence interaction between those of the Auger site as

$$H_U = \sum_{\mu_l < \nu_l} \sum_{\rho_l < \tau_l} U_{\mu_l \nu_l \rho_l \tau_l} a^+_{\mu_l} a^+_{\nu_l} a_{\tau_l} a_{\rho_l}. \qquad (7.1)$$

The diagrammatic method develops the time-dependent two-hole Green's function $G(t)$ in terms of local non-interacting time ordered one-body propagator

$$S_0(\alpha_l, \beta_l; t) = S_0^h(\alpha_l, \beta_l; t) - S_0^e(\alpha_l, \beta_l; -t) \quad (7.2)$$

where

$$S_0^h(\alpha_l, \beta_l; t) = -i\theta(t)\langle a^+_{\alpha_l}(t) a_{\beta_l} \rangle \qquad (7.3)$$

$$S_0^e(\beta_l, \alpha_l; -t) = -i\theta(-t)\langle a_{\beta_l} a^+_{\alpha_l}(t) \rangle \qquad (7.4)$$

and the average is taken over the non-interacting ground state $|\psi_0\rangle$, with energy $E_0$. In the BLA, one selects the series of ladder diagrams which are free of self-energy insertions and vertex corrections. In terms of the non-interacting two-hole Green's function $G_0$, the BLA leads to

$$G(\alpha_l \beta_l \beta'_l \alpha'_l; t) = G_0(\alpha_l \beta_l \beta'_l \alpha'_l; t)$$

$$-i \sum_{\mu_l < \nu_l} \sum_{\rho_l < \tau_l} U_{\mu_l \nu_l \rho_l \tau_l} \int_{-\infty}^{\infty} dt' \qquad (7.5)$$

$$G_0(\alpha_l \beta_l \tau_l \rho_l; t - t') G(\mu_l \nu_l \beta'_l \alpha'_l; t').$$

in frequency space, this becomes a linear algebraic system. Equation (7.5) is the exact solution for $n_h = 0$, and it is equivalent to keeping only those diagrams that remain in the closed-band limit. For closed bands, the one-body propagators of Equation (7.3) reduce to the $S_0^h$ part; therefore, lines that start as hole lines never go back in time and remain hole lines thoughout the diagrams. So, the order of times $t, t_1, t_2, ..., 0$ remains fixed (decreasing) despite the presence of the time-ordering operator T.

The convolution form of Equation (7.5) has furter important consequences. Mathematically, it is a Dyson equation in which the U matrix is an instantaneous self-energy. Therefore, it grants the <u>Herglotz property</u>: for any interaction strenght, G generates non-negative densities of states. The Herglotz property is a basic requirement for a sensible approximation, yet it is <u>not</u> easily obtained by diagrammatic approaches. The BLA [57,58] has been useful, among others, to interpret the line shape of Graphite [19].

*One-Step Model for early transition metals*

Experiments on early 3d transition metals, like Ti and Sc [64] could not be understood by the above theory. The maximum of the line shape was shifted by the interaction to lower binding energy, which is the contrary of what happens in closed band materials. Qualitatively the CS model could work if one admitted that U<0, and such an explanation has actually been proposed [65]; no known mechanism, however, produces an attractive U in the eV range. For almost empty bands one must formulate a new theory which is no simple extrapolation of the closed-band approach. Sarma [66] first suggested that the Auger line shape of Ti looks like some linear combination of the one-electron density of states and its convolution. This hint suggested to us that it was the two-step approximation that failed. The One-Step model [67] of the CVV Auger spectra has been put forth by Gunnarsson and Schönhammer; since the general formulation was too involved to use for the problem at hand, we simplified it as follows. Let $H_S$ represent Hamiltonian for the valence electrons without the core-hole, and $|\psi\rangle$ its ground state, that below we shall refer to as the *unrelaxed* one. We also need the Hamiltonian H' of the valence electrons in the presence of the core-hole potential, with its *relaxed* ground state $|\phi\rangle$. The Auger transitions are induced by the perturbation

$$H_A = \sum_{\alpha,\beta} M_{\alpha\beta} a_{\alpha l} a_{\beta l}; \quad (7,6)$$

this produces the Auger holes with matrix elements M in spin-orbitals denoted by Greek symbols; here operators are in the Heisenberg picture. For small $H_A$, the one-step expression for the Auger current of electrons with energy $\varepsilon_k$ is given by

$$J(\varepsilon_k) = \int_0^\infty dt_1 \int_0^\infty dt_2 f(t_1, t_2) \exp[i\varepsilon_k(t_1 - t_2)] \quad (7,7)$$

where

$$f(t_1, t_2) = \sum_{m,m'} \langle \psi | e^{i(H'+i\Gamma)} | m \rangle \langle m | H_A^+ e^{iH_S(t_1-t_2)}$$
$$H_A | m' \rangle \langle m' | e^{-i(H'-i\Gamma)} | \psi \rangle. \quad (7,8)$$

Here, $\Gamma$ is an operator which produces virtual Auger transitions and is approximated by a c-number; core and free-electron operators have already been averaged out; the m,m' summations run over a complete set of valence states. Since this results is still too involved for our purposes, we proposed [68, 69] that the complete set summations are largely exhausted by summing over just two <u>orthogonal</u> states, namely, *unrelaxed* and *relaxed* ground states $|\psi\rangle$ and $|\phi\rangle$. In this scheme, the Auger spectrum also has two main *relaxed* and *unrelaxed* contributions. The unrelaxed contribution arises from $|m\rangle = |m'\rangle = |\psi\rangle$ and can be expressed in terms of the two-hole Green's function $G(\omega)$. The relaxed contribution arises from $|m\rangle = |m'\rangle = |\phi\rangle$, and is proportional to

$$\langle \phi | a_\alpha^+ a_\beta^+ e^{iH_S(t_1-t_2)} a_{\beta'} a_{\alpha'} | \phi \rangle. \quad (7,9)$$

Experimentally, one can single out the relaxed contribution by fixing the photoelectron energy in an Auger-Photoelectron Coincidence Spectroscopy (APECS) experiment, [70-73] where the Auger electron is detected in coincidence with the photoelectron reponsible of the core hole creation. Fixing the photoelectron energy, the Auger electron measured in coincidence comes from the decay of a few dominant intermediate states in the presence of the core hole [74]. By a variational calculation Cini and Drchal [68-69] showed that, within a normalization constant, the following intuitive result holds,

$$|\phi\rangle \propto \sum_{\alpha l} a_{\alpha l}^+ a_F |\psi\rangle \quad (7,10)$$

where $a_{\alpha l}^+$ creates an electron in a localized spin-orbital at the Auger site and $a_F$ annihilates an electron in a suitable orbital at the Fermi level. In this way the screening cloud is represented by a single electron that has moved from the Fermi surface to the empty local states of the Auger site. Using Equations (7.10) and (7.9) one obtains an expression for the relaxed contribution to the Auger line shape involving the Fourier transform of the t>0 part of the 3-body Green's function

$$G_3(\alpha_l \beta_l \gamma_l \gamma'_l \beta'_l \alpha'_l; t)$$
$$= i \langle \psi | T \{ a_{\alpha_l}^+(t) a_{\beta_l}^+(t) a_{\gamma_l}(t) a_{\gamma'_l}^+ a_{\beta'_l} a_{\alpha'_l} \} | \psi \rangle. \quad (7.11)$$

## 8. TWO AND THREE-BODY PROPAGATORS

Since (7.11) is much harder to calculate than $G(\omega)$, recently we proposed [75] a simple approach in the spirit of the BLA where a new approximation called Core Approximation was introduced and tested with exact results from cluster calculations. This should allow to extend the analysis to several transition metals, giving at least a qualitative understanding of their spectra, which is currently a difficult task. In order to properly evaluate the results one should bear in mind that currently even for $G(\omega)$ we have reliable recipes only for $n_h$ less than $\approx 0.25$. This problem involves one more body and highly excited states of interacting systems; even the qualitative features of the solution are often quite an unsettled question.

*Ladder Approximation to the three-body propagator*

The Bare-Ladder Approximation to G($\omega$) (Sect. 7) is readily extended to the two-holes-one-electron $G_3(\alpha_l \beta_l \gamma_l \gamma'_l \beta'_l \alpha'_l; t)$; however, summing the Bare-Ladder series is essentially harder with three bodies than it was with two. In the two holes case, each interaction restarts the system with the two holes at the Auger site, and successive interactions yield independent factors in $\omega$ space; and that's why the ladder series is easily summed. With two holes and one electron, the situation is essentially more complex, as we may see for instance Figure 6a). When two bodies interact, the third one overtakes; therefore the diagram does not factor out at all.

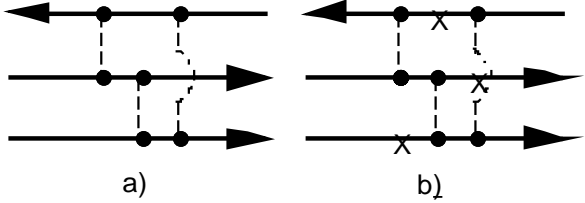

**Figure 6.** a) Typical BLA contribution to $G_3$. b) same as in a), with the formal three-body interaction.

Our approach to the problem is based on the idea that we can formally regard the two-body interaction as if it were a three-body one; the diagram 6a) then becomes 6b), with the fictitious X vertex which involves a summation over all sites. This is exactly true due to the identities, valid for all t'

$$S_0^h(\alpha_l, \beta_l; t) = i \sum_\gamma S_0^h(\alpha_l, \gamma; t-t') S_0^h(\gamma, \beta_l; t')$$
$$S_0^e(\alpha_l, \beta_l; t) = i \sum_\gamma S_0^e(\alpha_l, \gamma; t-t') S_0^e(\gamma, \beta_l; t') \quad (8.1)$$

The sums run over the complete set. In this way, introducing a fictitious X interaction vertex, along with the true one, every time-dependent diagram of the BLA is cast in a convolution form like (7.5), which is simplified by a Fourier transform. Sure, the infinite summations (one for each X interaction) are a high price to pay for that. Physically, in the spirit of the CST, we may expect that only the sites which are closest to the Auger site give important contributions to the summations, and we can actually work with a limited $\gamma$ set. Larger sets systematically lead to more precise results, at the cost of more computation. If you wish, this is a particularly efficient way to discretize the algorithm.

*Core-Approximation*

In the simplest approximation, the summation is limited to the local states $\gamma_l$. In this way the X vertex becomes local like the dot vertex; however some care is needed to preserve the correct t=0 limit, i.e., normalisation. Now we assume 0<t'<t and replace the exact identities (8.1) by the approximate ansatz

$$(-i)\langle a^+_{\alpha_l}(t) a_{\beta_l} \rangle$$
$$\approx \sum_{\gamma_l} R^+(\alpha_l, \gamma_l; t-t') \langle a^+_{\gamma_l}(t') a_{\beta_l} \rangle$$
$$(-i)\langle a_{\alpha_l}(t) a^+_{\beta_l} \rangle \quad (8.2)$$
$$\approx \sum_{\gamma_l} R^-(\alpha_l, \gamma_l; t-t') \langle a_{\gamma_l}(t') a^+_{\beta_l} \rangle;$$

the $R^\pm$ functions are computed by setting t'=0 in (8.2) and solving; so, the correct t=0 limit is granted. Equations (8.2) are correct in the limit of core states, when $S^{h,e}(\alpha_l, \beta_l; t)$ is diagonal in its indices and coincides with $R^\pm$; therefore we call this the *Core Approximation* (CA). The ansatz is also correct in the strong coupling case, when localised two-hole resonances develop. This is appealing, since the strong coupling case is the hard one, while at weak coupling practically every reasonable approach yields similar results. Ours is a physically motivated, simple approximation. We validated it by testing its results with exact results for its validation.

*Summing the Three-body ladder*

By the Core Approximation (CA) one can compute all kinds of ladder diagrams, to all orders. The partial sum of the series that one obtains in this way is referred to as *Core-Ladder-Approximation* (CLA); since both iteractions are local at the Auger site only local spin-orbitals appear in the equations and we can drop the notation $\alpha l, \beta l,\ldots$. At the first interaction time each of the three bodies can enter the X vertex, so there are 3 distinct contributions to the CLA at each order (figure 7) ; the contributions actually become 6 because each must be antisymmetrised with respect to the two incoming holes. In Figure 7 the black box represents $G_3$; the X on the top left corner represents the X vertex for an incoming particle or hole.

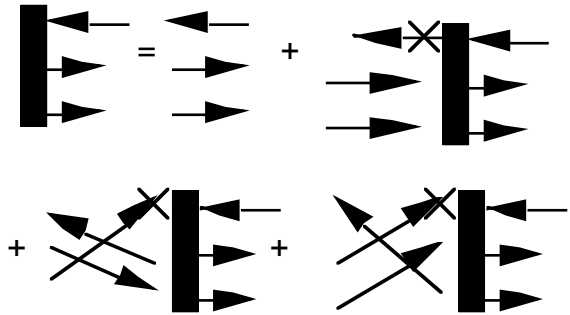

**Figure 7.** The Dyson-like equation for the local three-body Green's function $G_3$ in the CLA.

This is a closed system of equations for the local components of $G_3$. The first term on the rhs represents free propagation; the next introduces the effects of hole-hole interaction; the others come from electron-hole interactions and convey information on the screening effects due to the electronic cloud which forms as a response to the deep electron ionization.

The deep hole attracts a screening electron that can be directly involved in the Auger decay; locally, such processes leave the system with one hole in the final state. The presence of an one-body contribution in the Auger spectra from transition metals like Ti or Sc has been pointed out [66,68-69]. In the present scheme, this arises from the spin-diagonal components $G_3(\alpha\beta\gamma,\gamma'\,\beta'\,\alpha';t)$, (with $\sigma_{\alpha'}=\sigma_\alpha$ and so on); when the hole $\beta$ has the same z spin component as the electron, then, contracting $a_\beta^+(t)$ with $a_\gamma(t)$ and $a_{\beta'}$ with $a_{\gamma'}^+$ one obtains the extra term

$$G^{sp}(\alpha\beta\gamma,\gamma'\,\beta'\,\alpha';t) = \qquad (8.3)$$
$$-\langle a_\beta^+ a_\gamma\rangle\langle a_{\gamma'}^+ a_{\beta'}\rangle S(\alpha,\alpha';t),$$

where $S(\alpha,\alpha';t)$ stands for the time-ordered dressed one-body Green's function. Thus, neglecting small corrections [75], we write

$$G_3(\alpha\beta\gamma,\gamma'\,\beta'\,\alpha';\omega) = \qquad (8.4)$$
$$G^{CLA}(\alpha\beta\gamma,\gamma'\,\beta'\,\alpha';\omega) + G^{sp}(\alpha\beta\gamma,\gamma'\,\beta'\,\alpha';\omega).$$

Using the CLA, we get a conserving approximation for the proper self-energy. By solving Dyson's equation one can model XPS spectra from valence bands with low band filling; this is another field of application of our approach. To test the CLA, we compared its results with exact diagonalization data of a 5-atom model cluster with 2 orbitals for each atom. Accurate agreement between the CLA and exact results is obtained for high fillings and/or small U/W. Occupation numbers as low as <n>=0.72 and U/W ratios as high as 1 were considered. The CLA in such severe conditions is still in qualitative agreement with the results of the exact calculation. The performance of the CLA does not break down as quickly with increasing U as weak-coupling approaches tend to do, but remains fairly stable. The main reasons for this success are: 1) that electrons and holes are treated on equal footing, allowing a realistic treatment of screening 2) the theory becomes exact at weak coupling but also in the opposite limit of narrow bands (which is actually the core limit). 3)In all cases we find that the Herglotz property is fully preserved; this is a most valuable feature which is not easily obtained for approximate three-body propagators. Our theory explains the apparent *negative-U* behaviour: increasing the interaction U, the main peak shifts towards *lower* binding energies; this is a consequence of the interaction of the screening electron with the two Auger holes. We remark that a high enough $n_h$ is necessary to build up a localised screening cloud. This is why the *negative*-U behavior is observed in the early transition metals, but not in the late ones.

## 9. OUTLOOK

We presented a brief review of the state of advance of the theory of the Auger lineshape. Although the material selected is largely from our own work (thus reflecting our perspective of the subject) some general conclusions can be drawn and some trends for the future noted.

1) Auger spectroscopy is a valuable tool to probe excited states in solids, and to gain information on the "local " environment. As the experimental "resolution" improves steadily (especially by synchrotron radiation facilities) the experimentalists's goals become more ambitious in several directions. So we assist to the appearance of "old" type of experiments on new systems (magnetic multilayers, carbon compounds, manganites, controlled metal-oxide interfaces, etc. ), or to new experiments on "old"systems (coincidence, threshold, resonance experiments, two photon absorption, etc) and to the most exciting stuff, when both probing technique and sample are novel in some way.

2) Theory must keep up with these developments, to maintain Auger Spectroscopy "competitive" with other techniques. In fact, one can distinguish two major areas of progress: one related to the nature of the spectroscopical process itself , i.e. to include a more complex and accurate treatment of the de-excitation response within a One-Step-Model framework; this can be triggered by studies of traditional systems, but also by more sophisticated Auger experiments as mentioned above. The other direction is that of more complex systems, such as those with complex or poor screening, long-range interactions, complex magnetic/charge order, topological/ substitutional disorder, interplay between lattice and electronic degrees of freedom. With their ground and excited state novel behavior, they provide a complex, richer stage for the Auger decay, which represents the hardest challenge for the Auger open-band problem.

3) Computational approaches of spectroscopies such as optical absorption, XPS, EXAFS, etc. are at a more mature stage than those for solid state Auger spectroscopy ; for such processes there are sophisticated computer codes, which starting at the Density-Functional Theory , Hartree Fock, or similar level, allow to calculate the spectrum in a realistic way, including inelastic multiple scattering escape paths, surface geometry effects, temperature effects, etc.; in other words , by a robust numerical methodology, to investigate real systems with quantitative accuracy comparable to that of the experiment. Less has been made for solid state Auger spectroscopy in this respect, since the traditional difficulty in dealing with two- and three- body propagators at the *ab initio* level. However, due to significant computer advances, notable examples have recently appeared of abinitio treatments for two-body propagators. As a consequence, this is an area which holds promise of significant progress.

In short , we believe there are enough open and stimulating problems to provide Auger theoreticians with a very busy agenda for the next decade.